\documentclass[sort&compress]{elsarticle}
\usepackage{graphicx,amssymb,mathrsfs,amsfonts,amsmath}
\usepackage{comment}
\usepackage{enumerate}
\usepackage{subcaption}
\usepackage{color}

\usepackage{hyperref}
\hypersetup{
    pdfnewwindow=true,
    plainpages=false,
  }

\usepackage{tikz}

\usetikzlibrary{arrows.meta,decorations.pathmorphing,patterns,shadings}
\usepgflibrary{decorations.pathmorphing,shadings}
\newcommand{\be}{\begin{equation}}
\newcommand{\ee}{\end{equation}}
\def\bal#1\eal{\begin{align}#1\end{align}}
\def \d {{\rm d}}

\begin{document}
\title{Quantum fate of timelike naked singularity with scalar hair}

\author[1]{O. Sv\'{\i}tek}
  \ead{ota@matfyz.cz}
\author[1,2]{T. Tahamtan}
  \ead{tahamtan@utf.mff.cuni.cz}
\author[1]{Adamantia Zampeli}
  \ead{azampeli@phys.uoa.gr}
\address[1]{Institute of Theoretical Physics, Faculty of Mathematics and Physics, Charles University, V~Hole\v{s}ovi\v{c}k\'ach 2, 180~00 Prague 8, Czech Republic}
\address[2]{Institute of Physics, Faculty of Philosophy and Science, Silesian University in Opava, Bezru\v{c}ovo n\'{a}m. 13, CZ-74601 Opava, CZ}
\date{Received: date / Revised version: date}
\begin{abstract}
  We study the quantum fate of a naked curvature singularity sourced by a scalar field via several methods and compare the results obtained. The first method relies on relativistic quantum mechanics on a fixed background employing the Klein--Gordon and the Dirac equations for a static spacetime. We show that both the Klein--Gordon and the Dirac particles feel this singularity therefore this method does not provide its resolution. For comparison, we subsequently employ methods for quantizing the geometry itself. We selected the canonical quantization via conditional symmetries and as a last approach we use a maximal acceleration derivation in the covariant loop quantum gravity. In both of these approaches the singularity is resolved at the quantum level. We discuss these conflicting results bearing in mind that quantum particles probe classical geometry in the first approach while the last two methods quantize the geometry itself.
\end{abstract}
%
\maketitle
\section{Introduction}

Scalar fields are becoming increasingly relevant not only as a theoretical tool for explaining astrophysical phenomena (dark energy, inflation etc.) but the most important one of them has been detected --- the Higgs boson at the LHC. Regarding the static compact gravitating configurations of scalar field it is known from the Chase theorem \cite{Chase1970} (see \cite{Tafel:2011aa} for a recent generalization to a large class of potentials and a discussion of the role of energy conditions) that they do not possess a regular horizon (rather a naked singularity or a horizon coinciding with singularity appear). Due to this behavior of scalar field configurations we are going to investigate whether naked singularities persist on the quantum level as well.

{
  Here we will consider spacetime singularities to be defined by geodesic incompleteness. If they further give rise to diverging curvature scalars we call them curvature singularities --- this will be the case for the specific geometry under consideration.}

Curvature singularities generally appear in solutions of general relativity and show the limits of validity of this theory. When they are hidden beneath a horizon, they do not influence external observers and the situation is at least practically less serious (although the problem for the theory itself is not diminished). However, naked singularities represent a rather undesirable feature (motivating the Cosmic Censorship Hypothesis {--- here we consider its weak version prohibiting asymptotic observer from seeing singularity}) of a solution to the Einstein equations, especially if the associated matter content seems quite ordinary. Such a case is the one we study in this article. We investigate the possible resolution of a naked curvature singularity present in a recently derived solution belonging to the Robinson--Trautman family, minimally coupled to a free massless scalar field \cite{Tahamtan:2015sra} (a broader overview of the standard Robinson--Trautman solution with many references can be found there). {The Robinson--Trautman geometry is defined by the presence of a nonshearing, nontwisting and expanding null geodesic congruence. This family of spacetimes contains Schwarzschild or Vaidya black hole solutions but general members are lacking symmetries and evolve dynamically but do not include rotation. These spacetimes generically contain (exact) gravitational waves that carry away deviations from symmetry and vacuum subclass of these spacetimes settle down to the Schwarzschild solution asymptotically (proving its nonlinear stability within this subclass). The recent generalization \cite{Tahamtan:2015sra} of this family of spacetimes to include free massless scalar field source} contains a relatively wide range of special cases \cite{Tahamtan:2016fur} with some peculiar properties. For the purpose of this study we selected the static, spherically symmetric case which represents a parametric limit of the Janis--Newman--Winicour scalar field spacetime \cite{Janis:1968zz,Wyman:1981bd,Fisher:1948yn,Janis:1970kn}, which is asymptotically flat. In particular, this solution contains a timelike naked singularity (unlike the generic Janis--Newman--Winicour spacetime where the singularity is null) and the scalar field source satisfies standard energy conditions. As such this system represents an ideal model for investigating the peculiar classical behavior of compact static gravitating scalar field configurations on the quantum level.



It is generally expected that singularities will be remedied at a more fundamental level, in which both quantum and gravitational effects must be considered simultaneously. Our investigation concerns a possible resolution of the above mentioned naked timelike singularity, by employing three different quantum approaches and studying their implications at the semiclassical level. The first method is based on the pioneering work of Wald \cite{Wald:1980jn}, which was further developed by Horowitz and Marolf (HM) \cite{Horowitz:1995gi}. The main idea is to probe a classical timelike curvature singularity in static spacetime with quantum test particles obeying the Klein--Gordon equation. Later the method was applied in many specific geometries containing singularity \cite{Ishibashi:1999vw,Konkowski:2001th,Helliwell:2003tx,Konkowski:2011zz,Pitelli:2007jp,Pitelli:2008pa,Pitelli:2009kd,Tahamtan:2012sq,Tahamtan:2013vza}. In this approach, the singular character of the spacetime geometry is determined based on the number of self-adjoint extensions of an evolution operator. The evolution operator is extracted from the field equation selected for the analysis --- originally it was the Klein--Gordon equation, but the approach can be straightforwardly generalized to other field equations. The extended operator is then defined on a Hilbert space (usually an $\mathscr{L}_{2}$  space over a domain) covering the singularity position as well. If the self-adjoint extension is unique (so called essentially self-adjoint operator), it is said that the spacetime is quantum mechanically regular. This is connected to the fact that one can in general select a self-adjoint extension by demanding a specific boundary conditions for the eigenfunctions of the operator. However this cannot be applied in the singularity where we do not have any control over physics and therefore the extension should be unique automatically. This subsequently ensures a uniquely defined evolution for the wave-function thus mimicking a globally hyperbolic spacetime.

Although the above method is straightforward and clearly motivated it still treats the geometry classically. That is why we are going to present results from methods quantizing the geometry itself in order to provide more reliable answer and to compare the outcome with the previous method. A compelling argument why it is really necessary to pursue quantization of gravity instead of treating the spacetime geometry classically was presented by M. Bronstein already in 1936 \cite{Bronstein2012, Bronstein1936}. Since no generally accepted method for quantization of gravity exists we present results obtained in two radically different approaches, namely quantum geometrodynamics and covariant loop quantum gravity (CLQG) \cite{Rovelli:2014ssa}.

Quantum geometrodynamics is a canonical quantization of gravity with the three-metric taken as the configuration variable. The procedure relies on Dirac's quantization program since general relativity is a theory with constraints. Its central equation is the Wheeler--DeWitt equation coming from the Hamiltonian constraint which is then imposed in the operator version on a wave functions. In order to obtain the full content of the theory in the geometrodynamics description, no gauge fixing must be assumed at the classical level; this allows extra symmetries to appear which are lost otherwise. These extra symmetries, also known as conditional symmetries, preserve the isometries of the spacetime metric, thus it is natural to promote their corresponding charges to operators and impose them on the wave function together with the linear and quadratic constraints of the theory \cite{Kuchar:1982eb,Christodoulakis:2012eg,Christodoulakis:2013sya}. In \cite{Christodoulakis:2013xha}, the relation between the Lie point symmetries and the conditional symmetries of the minisuperspace was established which in the constant potential lapse parametrization coincide with the conditional symmetries on the phase space. For more details and applications see e.g. \cite{Zampeli:2015ojr,Dimakis:2017qcf,Karagiorgos:2017nta} and for a recent review \cite{sym10030070}. 

The last approach utilizes recent developments in CLQG, which build on previous results of the spinfoams approach \cite{Perez:2012wv}  and Loop Quantum Gravity (LQG) \cite{Thiemann:2007pyv}. Spinfoam approach attempts to use Feynman-style path-integral formulation where ``sum over geometries'' is considered in well-defined mathematical sense. It can provide natural tool for studying the dynamics of LQG from a covariant perspective. LQG represents canonical quantization of gravity where the geometry is encoded using connection (unlike in geometrodynamics). The connection is used to construct holonomies that play a central role in the formulation and the quantization leads to so-called spin network states. The geometry is then encoded in area and volume operators. The method we will use to study the singularity is based on the observation made in the realm of CLQG that there is a maximal acceleration in this theory \cite{Rovelli:2014ssa}. This upper bound appears in an analogous way to the minimal area in the original canonical Loop Quantum Gravity \cite{Rovelli:1994ge,Ashtekar:1996eg}. We derive a characteristic measure of acceleration in our spacetime and apply the upper bound yielding a resolution of our singularity. However, since we do not perform a complete derivation of the upper bound in our specific case this last approach should be treated with caution, although we follow the form of application suggested in \cite{Rovelli:2014ssa}. Nevertheless, the results regarding the geodesics and geodesic deviation that were necessary input for this part are interesting by themselves. 

In the rest of the paper we proceed as follows: in section \ref{robtrautscalar} we review the main properties of the classical spacetime of our interest, in section \ref{hm-method} we study the quantum properties at the semiclassical limit using the Horowitz--Marolf method, in section \ref{conditional} we perform the canonical quantization by the use of the conditional symmetries, in section \ref{covariant} we employ the CLQG ideas and finally in the last section we draw our conclusions.

\section{Static spacetime coupled to a scalar field} \label{robtrautscalar}
From now on, we focus on a particular case of the Robinson--Trautmann solution with a massless scalar field which was studied in \cite{Tahamtan:2016fur}. This specific spacetime is spherically-symmetric, static and contains a timelike naked singularity. The corresponding line element has the form 
\begin{eqnarray}\label{static-metric}
\d s^2 &=& \d {t}^2-\d {r}^2- \left({r}^2-{\chi_{0}^2}\right) \left( \d {\theta}^2+\sin^2{\theta}\,\d {\varphi}^2 \right)
\end{eqnarray}
and we have adopted the $(+,-,-,-)$ signature convention to retain the standard Newman--Penrose formalism choice. The matter is represented by a static massless scalar field in the following form 
\begin{equation}\label{static-field}
\Phi(r)=\frac{1}{\sqrt{2}}\ln{\left\{\frac{r -\chi_{0}}{r+ \chi_{0}}\right \}}\ .
\end{equation}
In the last paragraph of \ref{section-geodesic} it is shown that this spacetime is geodesically incomplete at $r=\chi_0$ and since the Ricci scalar and the Kretschmann invariant have the following form respectively
\begin{subequations}
\bal\label{RicciScalar}
&RicciSc=-\frac{2\,\chi_{0}^2}{\left({r}^2-\chi_{0}^2\right)^2}\ , \\
&Kretschmann=3(RicciSc)^2\ ,
\eal
\end{subequations}
one concludes that there is a curvature singularity at $r=\chi_0$ due to the above divergences. In the following we will usually shorten ``curvature singularity'' to purely ``singularity'' when we have this position in spacetime in mind.

One can easily observe that the singularity at $r=\chi_{0}$ (we consider only this one) is naked, either directly from the metric or by looking for marginally trapped surfaces. The singularity is pointlike and timelike. When ${r}\to \infty$, the scalar field vanishes and the metric (\ref{static-metric}) is asymptotically flat. The area of spherical surfaces $r=const., t=const.$ grows quadratically for values of the coordinate $r$ far from the central region (consistent with asymptotic flatness), while close to the singularity $r=\chi_{0}$ it grows only linearly. Further information about the geometry can be gained from geodesic motion and Penrose--Carter diagram (see \ref{section-geodesic}).

It is possible to shift the location of the singularity to zero by a coordinate transformation
\begin{equation}
\rho^2={r}^2-{\chi_{0}^2}\ ,
\end{equation}
which results in the metric
\begin{equation}\label{metric}
\d s^2 = \d {t}^2-\frac{{\rho}^2}{{\rho}^2+\chi_{0}^2}\d {\rho}^2-{\rho}^2\,\d \Omega^{2}\ .
\end{equation}
The newly introduced coordinate $\rho$ is a correct areal radius. However, for the subsequent calculations we retain the original form (\ref{static-metric}), since it leads to easier and more familiar expressions in both techniques analyzing the quantum aspects of the naked singularity at $r=\chi_{0}$.

\section{Self-adjoint extension method}\label{hm-method}
We are initially interested in probing the singular spacetime \eqref{static-metric} with test quantum particles --- specifically massless scalar and Dirac particles. To this end, we consider the method of self-adjoint extension which was introduced in \cite{Horowitz:1995gi} for probing singularities of the spacetime. Assume a static spacetime $\left( \mathcal{M},g_{\mu \nu }\right)\ $ with a timelike Killing vector field $\xi ^{\mu }$ and let $t$ denote the affine parameter along the Killing field and $\Sigma $\ denote a static spatial slice (with singular points removed). 
The Klein--Gordon equation for a scalar field can then be written in the following form
\begin{equation}\label{operator}
	\frac{\partial ^{2}\psi }{\partial t^{2}}=\sqrt{f}D^{i}\left( \sqrt{f}%
	D_{i}\psi \right) -fM^{2}\psi \equiv-A\psi \ ,
\end{equation}%
in which $f=\xi ^{\mu }\xi _{\mu }$ (using the selected signature of a spacetime metric) and $D_{i}$ is the spatial covariant
derivative on $\Sigma $ induced from the full spacetime covariant derivative, $M$ is the mass of the scalar field and $A$ is an operator defined on the Hilbert space $\mathscr{H}=\mathscr{L}_{2}\left( \Sigma \right)$ which is a space of square integrable functions on $\Sigma $. The operator $A$ is evidently real, positive and symmetric and therefore its self-adjoint extensions (covering the extension of Hilbert space to encompass the singular point) always exist. If this extension is unique, then $A$ is called essentially self-adjoint \cite{Horowitz:1995gi}. In order to analyze the essential self-adjointness, one has to consider the eigen-equation of the operator
\begin{equation}\label{essential}
	A\psi \pm i\psi =0\ ,
\end{equation}%
which will be called essentially self-adjoint if one of the two solutions of this equation (for each sign of the imaginary term) fails to be square integrable near the singularity. In this case, the operator can be unambiguously extended to the singularity and the corresponding wave functions are part of the Hilbert space. Such a system is then considered quantum mechanically regular. If $A$ is essentially self-adjoint for $M=0$, it is essentially self-adjoint for all $M>0$ as well \cite{Reed:1975uy}. 
\subsection{Klein--Gordon particle}
The Klein--Gordon equation for a massless scalar particle is given by
\begin{equation}
	\square \tilde{\psi} =g^{-1/2}\partial _{\mu }\left[ g^{1/2}g^{\mu \nu }\partial
	_{\nu }\right] \tilde{\psi} =0\ , 
\end{equation}%
which for the metric \eqref{static-metric}, it becomes
\begin{equation}\label{kg}
	\frac{\partial ^{2}\tilde{\psi} }{\partial t^{2}}=\left\{ \frac{\partial ^{2}}{\partial r^{2}}+\frac{2r}{r^{2}-\chi_{0}^{2}}\frac{\partial }{\partial r}+\right. \left. \frac{1}{r^{2}-\chi_{0}^{2}}\left(\frac{\partial ^{2}}{\partial \theta ^{2}}+\frac{1}{\sin ^{2}\theta }\frac{\partial ^{2}}{\partial \varphi ^{2}}+\frac{\cos\theta }{\sin \theta }\frac{\partial }{\partial \theta }\right) \right\} \tilde{\psi}\, .
\end{equation}
In analogy to equation \eqref{operator}, the spatial operator $A$ has the following form
\begin{equation}
	\emph{A}=-\left\{ \frac{\partial ^{2}}{\partial r^{2}}+\frac{2r}{r^{2}-\chi_{0}^{2}}\frac{\partial }{\partial r}+\right. \left.\frac{1}{r^{2}-\chi_{0}^{2}}\left(\frac{\partial ^{2}}{\partial \theta ^{2}}+\frac{1}{\sin ^{2}\theta }\frac{\partial ^{2}}{\partial \varphi ^{2}}+\frac{\cos\theta }{\sin \theta }\frac{\partial }{\partial \theta }\right) \right\}\, .
\end{equation}
Using a separation of variables, $\tilde{\psi}=e^{i\,\omega\,t}R\left( r\right) Y_{l}^{m}\left(
\theta ,\varphi \right) $, we obtain an equation for the radial function from equation (\ref{kg}). The left-hand side of the resulting equation is the radial part (which is the most important one for our analysis, since the remaining coordinates have compact ranges) of the operator $A$

\begin{equation}\label{radial}
	\frac{d^{2}R }{dr^{2}}+\frac{2r}{r^{2}-\chi_{0}^{2}}\frac{dR}{dr%
	}-\left( \frac{l\left( l+1\right) }{r^{2}-\chi_{0}^{2}}\right) R=-\omega^2R\ .
\end{equation}
The equation we have to study in order to decide about the essential self-adjointness is (\ref{essential}). So we have to deal with an ODE 
\begin{equation}\label{kg-last}
\frac{d^{2}\psi_{\pm} }{dr^{2}}+\frac{2r}{r^{2}-\chi_{0}^{2}}\frac{d \psi_{\pm} }{dr%
}-\left( \frac{l\left( l+1\right) }{r^{2}-\chi_{0}^{2}}\pm i\right) \psi_{\pm} =0\ .
\end{equation}
This is a Heun (singly) Confluent equation which is obtained from the general Heun equation containing four regular singularities through a confluence process; that is, a process where two singularities coalesce. This confluence procedure is performed by redefining parameters and taking limits resulting in a single (typically irregular) singularity \cite{ronveaux1995heun}. For the case of (\ref{kg-last}), there exist two regular singularities at $r=\pm \chi_{0}$ and one irregular at infinity. The solution for the above equation is expressed using Heun Confluent functions
\begin{eqnarray}
  \psi_{\pm}(r)&=&C_{1}\, HeunC\left(0,-\frac{1}{2},0,\mp\frac{i\chi_{0}^2}{4},\eta,\frac{r^2}{\chi_{0}^2}\right)\\
  &&+C_{2}\,r\,HeunC\left(0,\frac{1}{2},0,\mp\frac{i\chi_{0}^2}{4},\eta,\frac{r^2}{\chi_{0}^2}\right),\nonumber
\end{eqnarray}
where
\[\eta=\frac{1}{4}\left(\pm i\,\chi_{0}^2-l(l+1)+1\right)\ .\]
If we do not consider the subdominant (in the vicinity of singularity) term $\pm i$ in (\ref{kg-last}), the Heun Confluent functions simplify and the solution can be expressed in the following form  
\begin{equation}\label{kg-sol-noi}
\psi(r)=C_{1}\, P_{l}\left(\frac{r}{\chi_{0}}\right)+C_{2}\, Q_{l}\left(\frac{r}{\chi_{0}}\right)\ ,
\end{equation}
where $P,Q$ are the Legendre functions of the first and second kind respectively.

For the analysis of the square-integrability, it is worth to know the asymptotic behaviors of the above functions around the singular point $r=\chi_{0}$. The Legendre function $P_{l}\left(\frac{r}{\chi_{0}}\right)$ at $r=\chi_{0}$ is regular
\be
P_{l}(1)=1
\ee
and the Legendre function of the second kind, $Q_{l}\left(\frac{r}{\chi_{0}}\right)$, can be written as 
\begin{eqnarray}\label{LegendreQ}
Q_{l}\left(\frac{r}{\chi_{0}}\right)&=&\frac{1}{2} P_{l}\left(\frac{r}{\chi_{0}}\right)\ln\left[\frac{r+\chi_{0}}{r-\chi_{0}}\right]- \frac{2l-1}{l}P_{l-1}\left(\frac{r}{\chi_{0}}\right)-\ \cdots \ .
\end{eqnarray}
The square integrability of the solution (\ref{kg-sol-noi}) is checked by calculating a
squared norm in a proper functional space on each $t=const$
hypersurface $\Sigma $. We consider the Hilbert space $\mathscr{H}=\mathscr{L}_{2}\left(
\Sigma ,\mu \right)$, where $\mu $\ is a measure given by the spatial metric
volume element. It is straightforward to show that both solutions are square integrable at $r=\chi_{0}$ since the logarithmic divergence in (\ref{LegendreQ}) is compensated by the volume form ($[r^{2}-\chi_{0}^{2}]\d r$) to give a finite limit at $r=\chi_{0}$ for the integrand. One might be worried that, by removing the complex term from the equation, we have changed its nature too much. However, as shown in \cite{ronveaux1995heun} one of the solutions of Confluent Heun equation has (for the specific values of our parameters) logarithmic divergence --- as in the case of $Q_{l}$ --- and the other one is regular thus confirming that the result obtained via simplification holds for the full equation (\ref{kg-last}) as well.

Therefore, at this point, we conclude that the Klein--Gordon particle can see the singularity since the solution is square-integrable. 

\subsection{Massless Dirac particles}
In the case we have massless Dirac particles, the Newman--Penrose (NP) formalism \cite{Penrose:1986ca} is necessary to analyze the properties of the corresponding operator. The Chandrasekhar--Dirac (CD) equations \cite{Chandrasekhar:1985kt}, which represent a reformulation of the Dirac equation into the Newman--Penrose formalism, are suitable for this task and are given by
\begin{subequations}
\bal
	\left( D+\epsilon -\rho \right) F_{1}+\left( \bar{\delta}+\pi -\alpha
	\right) F_{2} &=0\ , \\
	\left( \nabla +\mu -\gamma \right) F_{2}+\left( \delta +\beta -\tau \right)
	F_{1} &=0\ ,  \label{CD} \\
	\left( D+\bar{\epsilon}-\bar{\rho}\right) G_{2}-\left( \delta +\bar{\pi}-%
	\bar{\alpha}\right) G_{1} &=0\ ,  \\
	\left( \nabla +\bar{\mu}-\bar{\gamma}\right) G_{1}-\left( \bar{\delta}+\bar{%
		\beta}-\bar{\tau}\right) G_{2} &=0\ , 
\eal
\end{subequations}
where $F_{1},F_{2},G_{1}$ and $G_{2}$ are the components of the Dirac wave
function (bispinor), $\epsilon$ ,$\rho$ ,$\pi$ ,$\alpha$ ,$\mu$ ,$\gamma$ ,$\beta $ and $\tau $ are the NP spin coefficients and the "bar" denotes a complex conjugation.
The null tetrad vectors for the metric (\ref{static-metric}) are defined by%
\begin{subequations}
\bal
	l^{a} =&\left( 1,1,0,0\right) , \\
	n^{a}=&\left( \frac{1}{2},-\frac{1}{2},0,0\right) ,  \\
	m^{a} =&\frac{1}{\sqrt{2(r^{2}-\chi_{0}^{2})}}\left( 0,0,1,\frac{i}{
		\sin \theta }\right) .  
\eal
\end{subequations}
The directional derivatives in the CD equations are given by $%
D=l^{a}\partial _{a},\nabla =n^{a}\partial _{a}$ and $\delta =m^{a}\partial
_{a}.$ To simplify the analysis, it is convenient to define auxiliary differential operators
\begin{subequations}
\bal\label{D-operators}
	\mathbf{D}_{0} &=D\ , \\
	\mathbf{D}_{0}^{\dagger } &=-2\nabla\ , \\
	\mathbf{L}_{0}^{\dagger } &=\sqrt{2(r^{2}-\chi_{0}^{2})} \,\delta \quad\mathrm{ and }\quad
	\mathbf{L}_{1}^{\dagger }=\mathbf{L}_{0}^{\dagger }+\frac{\cot \theta }{2}\ , \\
	\mathbf{L}_{0} &=\sqrt{2(r^{2}-\chi_{0}^{2})} \, \bar{\delta}\quad\mathrm{ and }\quad\mathbf{L}%
	_{1}=\mathbf{L}_{0}+\frac{\cot \theta }{2}\ .  
\eal
\end{subequations}
Evidently, the spatial parts of $\mathbf{D}_{0}$ and $\mathbf{D}_{0}^{\dagger }$ are purely radial operators, while $\mathbf{L}_{0,1}$ and $\mathbf{L}_{0,1}^{\dagger }$ are purely angular operators.

The nonzero spin coefficients for the metric (\ref{static-metric}) are given by
\begin{equation}
\rho=2\mu=-\frac{r}{{r^{2}-\chi_{0}^{2}}}\ ,\quad\beta =-\mathbf{\alpha }=\frac{1}{2%
		\sqrt{2}}\frac{\cot \theta }{\sqrt{r^{2}-\chi_{0}^{2}} }\ .
\end{equation}%
Substituting these nonzero spin coefficients and the definitions of the operators
(\ref{D-operators}) given above into the CD equations (\ref{CD}) leads to
\begin{subequations}
\bal
	\left( \mathbf{D}_{0}+\frac{r}{{r^{2}-\chi_{0}^{2}}}\right) F_{1}+\frac{1}{\sqrt{2(r^{2}-\chi_{0}^{2})}}%
	\mathbf{L}_{1}F_{2}=0\ , \\
	-\frac{1}{2}\left( \mathbf{D}_{0}^{\dagger }+\frac{r}{{r^{2}-\chi_{0}^{2}}}\right) F_{2}+\frac{1%
	}{\sqrt{2(r^{2}-\chi_{0}^{2})}}\mathbf{L}_{1}^{\dagger }F_{1}=0\ ,  \\
	\left( \mathbf{D}_{0}+\frac{r}{{r^{2}-\chi_{0}^{2}}}\right) G_{2}-\frac{1}{\sqrt{2(r^{2}-\chi_{0}^{2})}}%
	\mathbf{L}_{1}^{\dagger }G_{1}=0\ ,   \\
	\frac{1}{2}\left( \mathbf{D}_{0}^{\dagger }+\frac{r}{{r^{2}-\chi_{0}^{2}}}\right) G_{1}+\frac{1%
	}{\sqrt{2(r^{2}-\chi_{0}^{2})}}\mathbf{L}_{1}G_{2}=0\ .\label{CD-explicit}
\eal
\end{subequations}
To solve these CD equations, we assume a separable form of a solution
\begin{subequations}
\bal
	F_{1} &=f_{1}(r)Y_{1}(\theta )e^{i\left( kt+m\varphi \right) }\ , \\
	F_{2} &=f_{2}(r)Y_{2}(\theta )e^{i\left( kt+m\varphi \right) }\ ,  \label{separation-variables} \\
	G_{1} &=g_{1}(r)Y_{3}(\theta )e^{i\left( kt+m\varphi \right) }\ ,   \\
	G_{2} &=g_{2}(r)Y_{4}(\theta )e^{i\left( kt+m\varphi \right) }\ . 
\eal
\end{subequations}
Here $\left\{ f_{1},f_{2},g_{1},g_{2}\right\} $ and $\left\{
Y_{1},Y_{2},Y_{3},Y_{4}\right\} $ are functions of $r$ and $\theta $
respectively. Additionally, $m$ is the azimuthal quantum number and $k$ is the frequency of the
Dirac wave function, both are assumed to be real and positive. By substituting (\ref{separation-variables}) into (\ref{CD-explicit}) and using these
assumptions%
\begin{subequations}
\bal
	\mathrm{\ }f_{1}(r) &=g_{2}(r)\mathrm{ \ \ \ \ and \ \ \ }f_{2}(r)=g_{1}(r)%
	\mathrm{\ }, \\
	Y_{1}(\theta ) &=Y_{3}(\theta )\mathrm{ \ \ \ \ and \ \ \ }Y_{2}(\theta
	)=Y_{4}(\theta )\ ,
\eal
\end{subequations}
 we can reduce the system (\ref{CD-explicit}) into just two equations. The important radial parts of these two remaining Chandrasekhar--Dirac equations become
\begin{subequations}
\bal
	\left( \mathbf{D}_{0}+\frac{r}{{r^{2}-\chi_{0}^{2}}}\right) f_{1}\left( r\right) =\frac{\lambda 
	}{\sqrt{2(r^{2}-\chi_{0}^{2})}}f_{2}\left( r\right) , \label{CD-simplified}\\
	\frac{1}{2}\left( \mathbf{D}_{0}^{\dagger }+\frac{r}{{r^{2}-\chi_{0}^{2}}}\right) f_{2}\left(
	r\right) =\frac{\lambda }{\sqrt{2(r^{2}-\chi_{0}^{2})}}f_{1}\left( r\right) ,
\eal
\end{subequations}
where $\lambda $\ is a separation constant. For further simplification we introduce a new functions $\zeta _{1}, \zeta _{2}$ via
\begin{subequations}
\bal
	f_{1}\left( r\right)  =&\frac{\zeta _{1}\left( r\right) }{\sqrt{r^{2}-\chi_{0}^{2}}}\ , \\
	f_{2}\left( r\right)  =&\frac{\sqrt{2}\,\zeta _{2}\left( r\right) }{\sqrt{r^{2}-\chi_{0}^{2}}}\ ,
\eal
\end{subequations}
and the equations (\ref{CD-simplified}) transform into the following coupled system
\begin{subequations}
\bal
	\mathbf{D}_{0}\,\zeta _{1}(r)=\frac{\lambda}{\sqrt{r^{2}-\chi_{0}^{2}}}\,\zeta _{2}(r)\ , \\
	\mathbf{D}_{0}^{\dagger }\,\zeta _{2}(r)=\frac{\lambda}{\sqrt{r^{2}-\chi_{0}^{2}}}\,\zeta _{1}(r)\ .
\eal
\end{subequations}
or explicitly
\begin{subequations}
\bal
	\left( \frac{d}{dr}+ik\right) \zeta _{1}(r) &=\frac{\lambda}{\sqrt{r^{2}-\chi_{0}^{2}}}%
\,\zeta _{2}(r)\ , \\
	\left( \frac{d}{dr}-ik\right) \zeta _{2}(r) &=\frac{\lambda }{\sqrt{r^{2}-\chi_{0}^{2}}}%
\,\zeta _{1}(r)\ . 
\eal
\end{subequations}
In order to write the above equation in a more compact form we combine the
solutions in the following way,
\begin{subequations}
\bal
	\Xi_{+} =&\zeta _{1}+\zeta _{2}\ , \\
	\Xi_{-} =&\zeta _{2}-\zeta _{1}\ .
\eal
\end{subequations}
and square the operators to end up with a pair of one-dimensional Schr\"{o}dinger-like stationary equations with effective potentials,
\begin{subequations}
\bal
	\left( \frac{d^{2}}{dr^{2}}+k^{2}\right) \Xi_{\pm }=V_{\pm }\Xi_{\pm }\ , \\
	V_{\pm }=\frac{\lambda^{2}}{{r^{2}-\chi_{0}^{2}}}\mp \frac{r\lambda
	}{({r^{2}-\chi_{0}^{2}})^{\frac{3}{2}}}\ .
\eal
\end{subequations}
In analogy to the equation (\ref{operator}), the spatial operator $A$ for the massless
case is
\begin{equation}\label{A-Dirac}
	A=-\frac{d^{2}}{dr^{2}}+V_{\pm }\ ,
\end{equation}%
so the self-adjoint extension method (\ref{essential}) can now be applied to this operator which means that we have to analyze the solutions of
\begin{equation}
	\left( -\frac{d^{2}}{dr^{2}}+\left[ \frac{\lambda^{2}}{{r^{2}-\chi_{0}^{2}}}\mp \frac{r\lambda
	}{({r^{2}-\chi_{0}^{2}})^{\frac{3}{2}}}\right] \pm i\right)\psi _{\pm }=0\ .
\end{equation}%
To find the solutions of the above equation, we ignore the subdominant $\pm i$ term (which is negligible in the vicinity of singularity compared to other terms) and obtain 
\begin{equation}
  \psi _{\pm} =C_{1}\,\left(\pm 2\lambda \sqrt{r^{2}-\chi_{0}^{2}}+r\right)\left(\sqrt{r^{2}-\chi_{0}^{2}}+r\right)^{\mp \lambda}+C_{2}\left(\sqrt{r^{2}-\chi_{0}^{2}}+r\right)^{\pm\lambda}\ ,
\end{equation}%
in which $\lambda$ should be an integer. Obviously, when $r\rightarrow \chi_{0}$ (which is the singular point in our spacetime) the above two solutions are both finite and their Hilbert space norms near the singular point as well. Accordingly, the operator (\ref{A-Dirac}) is not essentially self-adjoint.
   
Consequently, the Dirac particle can see the existence of singularity on the spacetime, since all the solutions are square-integrable, a fact that renders the system quantum mechanically singular according to \cite{Horowitz:1995gi}. We are thus led to conclude that test particles of both bosonic and fermionic origin will ``feel'' the presence of the naked singularity of this specific background spacetime.

\section{Canonical Quantization via conditional symmetries}\label{conditional}
In a full quantum gravity theory, the geometry is dynamical and interacts with the matter fields, therefore the approaches which consider a background spacetime capture only low-energy features of the physical configuration and a non-perturbative approach is deemed necessary. Such an approach is given by the canonical quantization of the action for a Lagrangian in the minisuperspace form. 

In order to write an action principle for the spacetime under consideration, we first observe that the line element \eqref{static-metric} is of the general form 
\begin{equation}\label{generalmetric}
ds^2 = a^2 (r) dt^2 - \frac{N^2 (r)}{4 a^2(r)} dr^2 -b^2 (r) d\Omega^2\ ,
\end{equation}
where $a (r),b(r)$ are scale factors and $N(r)$ is the lapse function with $a=1, N=2, b= \sqrt{r^2 -\chi^2_0}$. Note, that we consider canonical evolution in the $r$-coordinate instead of the usual $t$-coordinate for cosmological minisuperspace models. Inserting \eqref{generalmetric} into the total action principle
\begin{equation}\label{total_action}
S_{tot} = S_{grav} +S_{mat}=\frac{1}{2}\int d^4 x \sqrt{g} \left(R+ g^{\mu\nu} \partial_\mu \phi \partial_\nu \phi \right)\ ,
\end{equation}
we observe that the Lagrangian function has the general minisuperspace form 
\be\label{general_lag}
\mathcal{L} = \frac{1}{2N(r)} \mathcal{G}_{\alpha \beta}  (q^\alpha (r))\dot{q}^\alpha (r) \dot{q}^\beta (r) - N (r) \mathcal{V} (q^\alpha (r))\ ,
\ee
where $r$ is the radial coordinate and an independent variable, $q^\alpha (r)$ are the dependent variables denoting the gravitational and matter degrees of freedom $a(r), b(r), \phi (r)$ with $\mathcal{G}_{\alpha \beta}$ being metric on the configuration space of the dependent variables, known as a supermetric and $\mathcal{V}(q)$ is the superpotential, which usually contains terms related to the curvature of the hypersurface which foliates the spacetime. From the form of the Lagrangian \eqref{general_lag}, it is evident that this model is reparametrisation invariant with respect to the radial coordinate, a fact consistent with the static nature of the spacetime.

If one considers the line element \eqref{generalmetric}, we find that the Lagrangian, after a reparametrisation so that the superpotential becomes constant \cite{Christodoulakis:2012eg}, $\mathcal{V} (q) = 1$, takes the form 
\begin{equation}\label{lag}
\mathcal{L}= -N - \frac{8aba'b'}{N}-\frac{4a^2b'^2}{N}+\frac{2 a^2 b^2 \phi'^2}{N}\ ,
\end{equation}
where $' \equiv \frac{d}{dr}$. {The convenience of the constant potential parametrization (which is not a gauge choice) will become evident shortly.} It is not difficult to see, by inspection of the kinetic term, that the supermetric has the form\footnote{The form of the supermetric is the same as in the case of Kantowski-Sachs spacetime coupled to a massless scalar field, see \cite{Zampeli:2015ojr}.}
\begin{eqnarray}
\mathcal{G}_{\alpha \beta} =\left(
\begin{array}{cccc}
0 & -8ab &0\\
-8ab & -8 a^2&0\\
0 & 0 & 4 a^2 b^2\\
\end{array}
\right)\ .\label{supermetric}
\end{eqnarray}
The usual approach to canonical quantization is to find the dynamical equation (Wheeler--DeWitt) by imposing the classical Hamiltonian constraint on the wave function, which for the model under consideration is
\begin{equation} \label{wdw}
  \hat{\mathcal{H}} \Psi = \frac{1}{32 a^2 b^2} \left( \left( -1 + 32 a^2 b^2\right) -\right. \left. 2 \left( 2 \partial_{\phi \phi} - b \partial_b + a \left( \partial_a - 2b \partial_{ab} + a \partial_{aa} \right) \right) \right) \Psi = 0.
\end{equation}
{To arrive at this form of the equation, the following have been implemented: the canonical variables are the $q^{\alpha}$'s and their conjugate momenta which have been promoted to operators by the rule $q^\alpha \rightarrow \hat{q}^\alpha$ and $p_\alpha \rightarrow \hat{p}_\alpha = -i \frac{\partial}{\partial q^\alpha}$ together with the Poisson brackets becoming commutators, $\{.,.\} \rightarrow [.,.]$. The operator ordering in the Hamiltonian constraint is such that the kinetic term is the conformal Laplacian and the operators are Hermitian under a proper choice of measure \cite{Christodoulakis:2012eg}.}
Here, we consider the additional symmetries of the superspace, which is conformally flat, as operators acting on the wave function. This step is justified for several reasons, one being that they are related to the outer automorphisms group which leaves the geometry invariant under general coordinate transformations. The Hamiltonian constraint, as well as the diffeomorphisms constraints, which in the simplified case of the minisuperspace models, vanish identically forms the inner automorphisms \cite{Christodoulakis:2001mg,Zampeli2019}.

These symmetries are given by three Killing vector fields 
\begin{eqnarray}
\xi_1 =-a\partial_a+ b\partial_b\ , \quad
\xi_2 =-a\phi \partial_a + b\phi \partial_b+2 \ln a \partial_\phi\ , \quad 
\xi_3 =\partial_\phi\ .
\end{eqnarray}
{which in a general parametrization with $\mathcal{V} \neq const$ are conformal Killing vector fields. The advantage of taking it as constant is that} on the phase space, we can construct linear to the momenta quantities of the form $Q_i = \xi_i^\alpha p_\alpha$, which are conserved by virtue of the constraint. In our case these are 
\begin{subequations}\label{conservedclassical}
\bal
Q_1 &=-\frac{8a b^2 {a}'}{N} = \kappa_1 \label{conserved_Q1}\ , \\
Q_2 &=-\frac{8 a b^2 \left( \phi {a}' -a {\phi}' \ln a \right)}{N} =\kappa_2 \label{conserved_Q2}\ ,\\
Q_3 & = \frac{4 a^2 b^2 {\phi}'}{N}=\kappa_3 \label{conserved_Q3}\ ,
\eal
\end{subequations}
where we have replaced the canonical momenta as $p_{a}=-\frac{8a b b'}{N}$,$\ p_{b}=-\frac{8a(ba'+ab')}{N}$,\\$\ p_{\phi}=\frac{4a^2 b^2 \phi'}{N}$. For the gravitational system of our interest, the constants take the values 
\begin{equation}\label{valuesconsts}
\kappa_1=\kappa_2=0\ , \quad \kappa_3= 2\sqrt{2} \chi_0\ .
\end{equation} 
Under the demands that general covariance is preserved at the quantum level, thus the Dirac algebra of the $\hat{Q}_i$'s be isomorphic to their classical commutation relations,\footnote{The commutation relation for the first integral quantities come from the Lie bracket algebra of the corresponding Killing vector fields $[\xi_1,\xi_2]=- 2 \xi_3, [\xi_2,\xi_3]= -\xi_1$.} together with the selection of a proper measure to ensure the Hermiticity of the operators \cite{Christodoulakis:1984gp}, lead to the selection rule $c_{ij}^k \kappa_k =0$ {for the quantum charges which correspond to \eqref{conservedclassical} so that not all of these operators can be imposed simultaneously \cite{Christodoulakis:2012eg}. In this case, we have two possible operator subalgebras $\{ \hat{Q}_1, \hat{Q}_3\}$ and $\{ \hat{Q}_2\}$.}

In the following, we focus our study on the case of the two-dimensional subalgebra $\{ \hat{Q}_1, \hat{Q}_3\}$ and solve the following equations
\begin{subequations}
\bal
\hat{Q}_1 \Psi & = i \left(  -b \partial_b + a \partial_a  \right) \Psi = \kappa_1 \Psi\ , \\
\hat{Q}_3 \Psi & = -i \partial_\phi \Psi = \kappa_3 \Psi\ ,
\eal
\end{subequations}
together with the constraint equation (\ref{wdw}). The solution of this system is given in terms of the spherical Bessel functions
\begin{eqnarray}\label{Psi-origin}
&\Psi(a,b,\phi) = e^{i 2 \phi \chi_0} \left( A_1 J_{\lambda} (4ab) +A_2 Y_{\lambda} (4ab)\right),
\end{eqnarray}
where $\lambda = \frac{1}{2} (-1 + \sqrt{3-64\chi_0^2})$.

To get a rough idea about the consequences that the above wave function has for the fate of the singularity, one can consider probability distribution on the superspace (leaving out the question of normalization of the states). Specifically, one shall consider probability density given by states of the form (\ref{Psi-origin}) including the measure coming from the metric (\ref{supermetric})
\begin{equation}\label{prob-density}
	\mathfrak{p}=|\Psi|^{2}\, 16\,a^{2}b^{2}\ .
\end{equation}
\begin{figure}[t]
	\centering 
	\begin{subfigure}[b]{0.49\textwidth}
	\centering 
	\includegraphics[width=\textwidth]{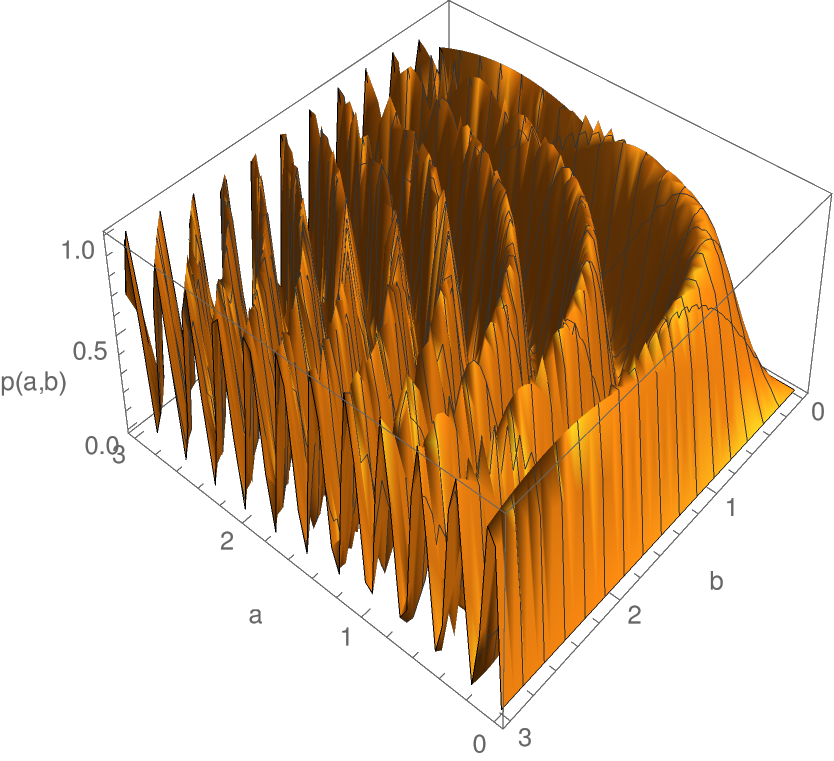}
	\subcaption{$A_{1}=1,A_{2}=0 \Rightarrow$ $\mathfrak{p}=J_{\lambda}^2 (4ab) 16 a^2 b^2$ }
	\label{BesselJ}
		\end{subfigure}
	\begin{subfigure}[b]{0.49\textwidth}
	\centering 
	\includegraphics[width=\textwidth]{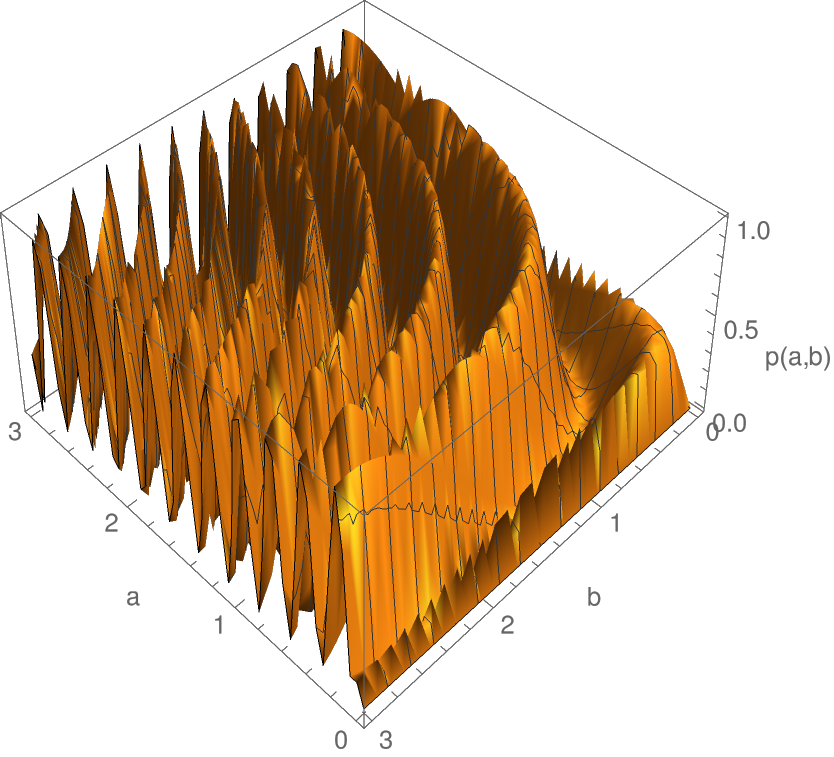}
	\subcaption{$A_{1}=0,A_{2}=1 \Rightarrow$ $\mathfrak{p}=Y_{\lambda}^2 (4ab) 16 a^2 b^2$}
	\label{BesselY}
	\end{subfigure}
	\caption{Plots of the probability density $\mathfrak{p}=|\Psi|^{2}\, 16\,a^{2}b^{2}$ in the intervals $a \in [0,3]$ and $b \in [0,3]$ and with $\chi_{0}=\frac{\sqrt{2.99}}{8}$ for the wave function \eqref{Psi-origin} for two cases --- plot (a) for $A_{1}=1,A_{2}=0$ and plot (b) for $A_{1}=0,A_{2}=1$ (index of both Bessel functions is  $\lambda = \frac{1}{2} (-1 + \sqrt{3-64\chi_0^2})$). It can be seen that when $a$ or $b$ approach zero, the probability density in both cases vanishes.}
\end{figure}
{As one can observe from the plots of the probability density \eqref{prob-density} on the plane of variables $a,b$ for two choices of integration constants --- first one for $A_{1}=1,A_{2}=0$ corresponding to BesselJ function only (Figure \ref{BesselJ}) and the second one for $A_{1}=0,A_{2}=1$ corresponding to BesselY function only (Figure \ref{BesselY}) --- the probability density of the system to be in the state corresponding to either $a=0$ or $b=0$ is highly suppressed. This means that quantum geometry avoids spacetimes containing singularity in our case. Such interpretation is in accord with the Hartle's criterion for making prediction in quantum cosmology \cite{Hartle_1987}.} We have selected a value of the parameter $\chi_0$ close to the value $\frac{\sqrt{3}}{8}$ beyond which the index of the Bessel functions becomes complex. Since the physical singularity for the metric (\ref{generalmetric}) appears only for $b=0$ (as one can easily derive from the corresponding Kretschmann scalar) on the quantum level this situation seems to be avoided.

Another possible approach to derive physical consequences from the wave function employs the Bohmian interpretation \cite{Bohm:1951xw,Bohm:1951xx}. This approach is especially useful in quantum cosmology since it enables one to define quantum paths on the configuration space through the guiding equations 
\begin{equation}\label{guidance_eq}
\frac{\partial \mathcal{L}}{\partial \dot{q}^\alpha} = \frac{\partial S}{\partial q^\alpha}\ .
\end{equation}
These are defined through the identification $p_\alpha \equiv \partial_\alpha S$ and $S(q)$ is the phase in the polar form expression of the wave function, $\Psi (q)= \Omega (q) e^{iS(q)}$. As in the case of the Schr\"{o}dinger equation, inserting this form in the generic Wheeler-DeWitt equation
\begin{equation}\label{hamiltonianpoint}
\hat{\mathcal{H}}\Psi \equiv \left(-\frac{1}{2 \sqrt{\mathcal{G}}} \partial_\mu \mathcal{G}^{\mu\nu} \partial_\nu - \frac{d-2}{8(d-1)} \mathcal{R} + 1\right) \Psi =0\ ,
\end{equation}
where $d$ is the dimension of the configuration space and $\mathcal{R}$ is its corresponding Ricci scalar, we obtain a modified Hamilton-Jacobi equation
\begin{equation}\label{modified_HJ} 
\frac{1}{2} \mathcal{G}^{\alpha \beta} \partial_\alpha S \partial_\beta S -\frac{1}{2} \frac{\Box \Omega}{\Omega} +1 =0\ .
\end{equation}
This, together with the guiding equation \eqref{guidance_eq}, determines the behaviour of the quantum system.
Equation (\ref{modified_HJ}) contains an additional potential term compared to its classical analogue. If this term is nonzero, the solution of (\ref{guidance_eq}) will be different from the classical one, while when it vanishes we should recover the classical spacetime. 

For our particular case, we will consider the approximation $A_1 \rightarrow 0$ (thus concentrating on the Bessel function diverging at the origin) for small ($sm$) and large ($la$) arguments of the spherical Bessel function to bring the wave function in polar form. Thus we obtain respectively
\begin{subequations}
\bal
	&\Psi_{sm} = C_1 (a b)^{-\frac{1}{2} -\frac{1}{2} \sqrt{3-64 \chi_0^2}}  e^{i 2\chi_0 \phi}\ ,\label{Psi-small} \\
	& \Psi_{la} = C_2 \frac{\sin \left(-4ab + \frac{\pi}{4} \left(-1 + \sqrt{3-64\chi_0^2}\right) \right)}{4 ab} e^{i 2\chi_0 \phi}\ .
\eal
\end{subequations}
The above cases give us two different solutions both of which differ from the classical solution of the system (\ref{conservedclassical}), since the quantum potential does not vanish for neither of them. We assume two subcases for the small arguments ($sm$), corresponding to negative or positive value of the quantity under the square root in (\ref{Psi-small}) (since the function $S (a,b, \phi)$ differs accordingly). Then the spacetime elements we obtain are
\begin{subequations}
\bal
	&ds^2 =  dt^2 - dr^2 - \lambda_1 d \theta^2 -\sin^2 \theta d\phi^2\ , \label{small_arg} \\
	&ds^2 = \lambda_2 dt^2 -\lambda_2  r^2 dr^2 - r^2 (d\theta^2 + \sin^2 \theta d\phi^2)\ ,
\eal
\end{subequations}
where $\lambda_1, \lambda_2$ are essential constants which characterize the geometry of spacetimes. 

For large arguments, the solution is again (\ref{small_arg}). The curvature scalars of these line elements inform us that it is only for the range $-\frac{\sqrt{3}}{8} < \chi_0 < \frac{\sqrt{3}}{8}$ of the constant $\chi_0$ for the small arguments and any range for the large arguments that the singularity vanishes from the semiclassical line element.

The above results show that either directly by using the probability density, which vanishes for either $a$ or $b$ going to zero, or by constructing effective geometries via Bohmian approach we arrive at the conclusion that the singularity is removed as a place of both curvature divergence and geodesic incompleteness. Since the singularity is removed, the quantum version of spacetime is in agreement with the Cosmic Censorship Hypothesis.

\section{Covariant Loop Quantum Gravity}\label{covariant}

To provide an alternative derivation in the realm of quantum gravity we will turn to CLQG \cite{Rovelli:2014ssa}. When the static spacetime \eqref{static-metric} possesses a horizon covering the central spacelike singularity one can use the Loop Quantum Cosmology (LQC) method since the spacetime below the horizon (which is no longer static) can often be mapped onto some symmetric cosmological model (e.g. Kantowski-Sachs) whose singularities are generally resolved in LQC. Our spacetime however contains a naked timelike singularity so we cannot use this trick. Instead we can apply the recent discovery on the level of the CLQG \cite{Rovelli:2013osa} that quite generally the singularities are resolved due to the upper bound for the acceleration of observers arising at the quantum level. This derivation is based on considering the Rindler observers but has been  later applied to cosmology with the characteristic acceleration being the mutual acceleration of nearby comoving observers. For the spacetime in question, we consider essentially the same quantity, a relative acceleration with respect to a radial geodesic, as given by the geodesic deviation equation. This also measures the tidal forces acting upon an object approaching the singularity.

The important properties of geodesics in the spacetime of interest are summarized in the \ref{section-geodesic}. The four velocity of a radial geodesic (considered in the equatorial plane for simplicity) is described by
\begin{equation}
  u^{\mu}\partial_{{\mu}}=\sqrt{(u^{r})^{2}+1}\,\partial_{t}+u^{r}\,\partial_{r}\ ,
\end{equation}
with the radial velocity $u^{r}$ being a constant. The deviation vector is considered in the form ${\mathbf{\delta}}=\delta^{\phi}\, \partial_{\phi}$. The geodesic deviation equation then assumes the following form
\begin{equation}\label{geodesic-deviation}
  \frac{D^{2}{\delta^{\alpha}}}{d\tau^{2}}=-R^{\alpha}{}_{\beta\gamma\sigma}u^{\beta}\delta^{\gamma}u^{\sigma}=-\frac{\chi_{0}^{2} (u^{r})^{2}}{(r^{2}-\chi_{0}^{2})^{2}} \delta^{\alpha}\ .
\end{equation}
Evidently, the tidal force grows unbounded when approaching the singularity even though the radial geodesic observer is not accelerated with respect to the asymptotic observer (see the end of \ref{section-geodesic}). As a measure of the acceleration, we will use the invariant norm of (\ref{geodesic-deviation}) with respect to a unit separation
\begin{equation}\label{acceleration}
 a=\frac{\chi_{0}^{2} (u^{r})^{2}}{(r^{2}-\chi_{0}^{2})^{3/2}}\ .
\end{equation}
According to \cite{Rovelli:2013osa}, the acceleration is bounded by a maximum value $a_{max}\sim \sqrt{\frac{1}{8\pi G \hslash}}$ (in nongeometric units). This result is moreover derived in a fully covariant theory unlike previous upper bounds to the acceleration \cite{Magueijo:2001cr}. Inspecting (\ref{acceleration}), one immediately sees that the upper bound to the acceleration means that the divergent factor ${(r^{2}-\chi_{0}^{2})}^{-1}$ appearing in the curvature scalars (\ref{RicciScalar}) is also bounded and therefore the singularity (as a place of diverging curvature) is resolved at the level of CLQG. The question of geodesic completeness is more subtle since we do not have access to reconstructed spacetime geometry in this quantization approach. However, since all the relevant operators have bounded spectrum it points to an effective classical geometry which should be geodesically complete. Additionally, the tidal forces are bounded and an object can in principle survive the fall into the singularity (or to the region where the curvature singularity appears classically). However, the bound is extremely large so it is hard to imagine any realistic object that would not be crushed. 

So one can conclude that the critical behaviour of General Relativity (its breakdown at the position of singularity) is cured at the quantum level (infinities are removed), but the practical result of approaching the singularity (destruction of an extended object) remains effectively the same. To interpret the result from the point of view of Cosmic Censorship Hypothesis, we would need to understand the effective geometry. However, the bound on acceleration (\ref{acceleration}) means that the only nontrivial metric function $(r^{2}-\chi_{0}^{2})$ should not vanish. Based on this one can argue that any potential effective geometry should contain a modification preventing this metric function from attaining zero and negative values thus removing obstruction to continuation of geodesics through the point in discussion (see the last paragraph of \ref{section-geodesic} for the discussion of incompleteness of the original classical geometry). In this way, the Cosmic Censorship Hypothesis is saved since the singular nature of spacetime was completely removed and therefore all observers do not have any causal contact with singularity.

In the original paper \cite{Rovelli:2013osa} the upper bound on acceleration is derived based on the quantization of timelike two-dimensional surface related to the accelerated trajectory. Since the area of this surface depends on the acceleration its quantization leads to bounds on both. We show in \ref{appendixb} that in our case there is also an explicit relation between the above derived acceleration and a timelike two-surface naturally associated to the pair of geodesics under consideration.

Although we used the results of \cite{Rovelli:2013osa} in the way suggested therein we did not completely rederive the whole quantization procedure (introducing operators etc.). That is why the results of this section should be regarded as secondary to the canonical quantization result of Section \ref{conditional}. However, they still provide useful confirmation from a completely different perspective.

\section{Conclusion and final remarks}
In this article, we studied the quantization of a spacetime with naked singularities by employing three different approaches: one is the HM method for the study of classical timelike curvature singularities, the second the canonical quantization via conditional symmetries and the last the examination of the existence of a maximal acceleration in the realm of the CLQG.

In the context of the first methodology, we have shown that for both the Klein--Gordon particle and the Dirac particle, all solutions of (\ref{essential}) are square integrable. This means that the corresponding operators in both cases are not essentially self-adjoint and the problem is quantum mechanically singular. Therefore, the quantum probes still see the singularity in this case. This is not true in the case of the second approach, where we found that under reasonable assumptions we can find quantum corrections to the geometry which resolve the singularity. This was shown by employing both the Bohmian approach of quantum theory at the semiclassical level as well as by examining the probability density distribution on the superspace. Finally, in the case of the CLQG the maximal acceleration existence provides the means to effectively remove the singularity as demonstrated above. However, one should be cautious regarding this argument since our case provides an indication rather than a complete proof for the general case (see the last paragraph of Section \ref{covariant} as well). At the same time, the implications for an observer approaching the position of a now resolved singularity seem catastrophic even in this quantum picture because the upper bound on tidal forces is extremely large. Nevertheless, both of the spacetime quantization methods save the Cosmic Censorship Hypothesis by effectively removing the singularity.

It is evident that the spacetime quantization approaches yield results contradicting the quantum particle approach. Since these methods are based on quantum description of spacetime, one should give them preference over the quantum particle approach where the spacetime itself is classical and only the probes are quantum. Our results also support the argument that it is in the realm of a quantum gravity theory where the (naked) singularities are reliably resolved. Further comparison between these quantization methods by employing other singular spacetimes, e.g. the recently studied Janis--Newman--Winicour solution \cite{Svitek:2014fga}, can shed more light on the singular behavior of compact scalar field configurations and appropriateness of each approach to the quantization of a gravitational systems.

Insight could also be provided by the use of quantum field theory on a curved background with the inclusion of semiclassical backreaction effects. This would fit in-between the approaches presented here, since using quantum probe field is certainly closer to a realistic scenario than relying only on the quantum mechanical particles. On the other hand, these approaches should be superseeded by a spacetime quantization. The presented quantum gravity approaches relied on highly symmetric nature of the geometry under consideration and should be pursued in more generic situations, e.g. using canonical quantization or CLQG to perform full spacetime quantization going beyond spherically symmetric models. {Even though the minisuperspace approximation does not capture the full dynamical content of a quantum theory of gravity, since most of the degrees of freedom are ``frozen", it is still possible to derive essential insight for the features of the full theory of quantum gravity}. However, currently only initial tentative steps are being made in this direction {and no final answer exists}.

\section*{Acknowledgments}
This work was supported by the research grant GA\v{C}R 17-16287S and the INTER-EXCELLENCE project No. LTI17018 that supports the collaboration between the Silesian University in Opava and the Astronomical Institute in Prague. We also acknowledge the endorsement of the Albert Einstein Center for Gravitation and Astrophysics, Czech Republic.

\appendix
\section{Geodesic equation}\label{section-geodesic}
In this appendix we study the trajectory for a test particle moving on a timelike geodesic giving us further insight into the nature of the singularity under investigation. The simplest approach is to use the variational principle or the Euler--Lagrange equations for timelike geodesics. The Lagrangian reduces to kinetic part for the free particle under the influence of gravity only (which is encoded in geometry via metric) and has the following form
\begin{eqnarray}\label{metric-geod}
2L=g_{\mu\nu}\,\dot{x}^{\mu}\dot{x}^{\nu}&=&\dot{t}^{2}-\dot{r}^{2}- \left({r}^2-{\chi_{0}^2}\right)\,\left(\dot {\theta}^2+\sin^2{\theta}\,\dot {\varphi}^2\right)\ ,
\end{eqnarray}
where the dot denotes a derivative with respect to the proper time $\tau$. The Euler--Lagrange equations 
\begin{equation}
\frac{\partial L}{\partial x^{\mu}}-\frac{d}{d\tau}\left(\frac{\partial L}{\partial\dot{x}^{\mu}}\right)=0\ ,
\end{equation}
give us two conserved quantities, namely the energy $(E)$ and the angular momentum $(l)$
\begin{subequations}\label{El}
\bal
& \frac{d}{d\tau}\left(\frac{\partial L}{\partial\dot{t}}\right)=0 \Rightarrow {\dot{t}}=E\ ,\label{Elb} \\
& \frac{d}{d\tau}\left(\frac{L}{\partial\dot{\varphi}}\right)=0 \Rightarrow \dot{\varphi}=\frac{l}{\left({r}^2-{\chi_{0}^2}\right)\sin^2{\theta}}\ .
\eal
\end{subequations}
We consider motion in the equatorial plane $\theta=\frac{\pi}{2}$. Substituting (\ref{El}) in (\ref{metric-geod}), we obtain
\begin{equation}\label{radialgeo}
E^2-\dot{r}^{2}-\frac{l^2}{\left({r}^2-{\chi_{0}^2}\right)}=1\ .
\end{equation}
\begin{figure}
	\centering 
	\includegraphics[scale=0.6,bb=60 400 400 750]{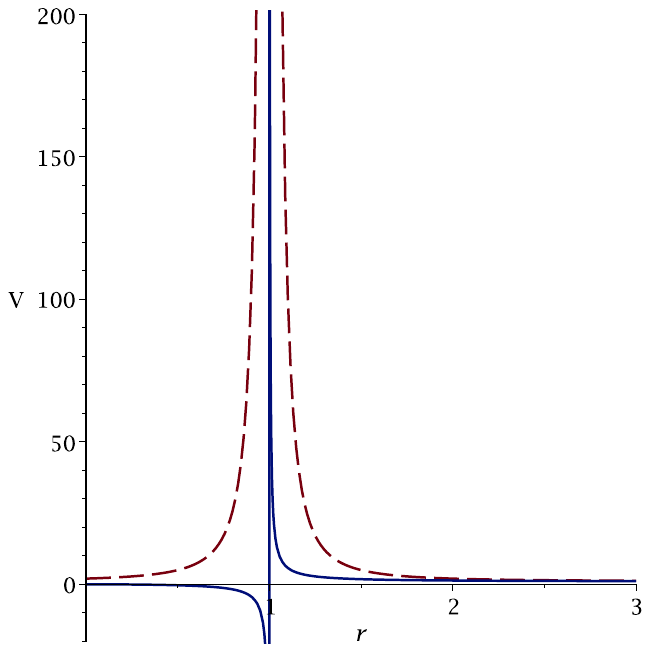}
	\caption{Plot of the effective potential  ${V}_{eff}=1+\frac{l^2}{\left({r}^2-{\chi_{0}^2}\right)}$                   for angular momentum $l=1$ and $\chi_{0}=1$ (solid line) compared with standard centrifugal barrier (dashed line) with respect to center shifted to position $r=\chi_{0}$ instead of $r=0$.}
	\label{plot}
\end{figure}
For a qualitative analysis of geodesics, we employ the standard effective potential method. Then we can write the equation of the radial velocity (\ref{radialgeo}) in the form
\begin{eqnarray}
\dot{r}^{2}+{V}_{eff}=E^2\ ,  \quad {V}_{eff}=1+\frac{l^2}{\left({r}^2-{\chi_{0}^2}\right)}\ .\label{effective}
\end{eqnarray}

The effective potential is plotted in Figure \ref{plot}. It is evidently repulsive and acting similarly to a centrifugal barrier (in a flat space) and in fact its origin is similar. Compared to the centrifugal barrier, it allows the particle to travel closer to the origin (at $r=\chi_{0}$). Note that for vanishing angular momentum $l$, the radial velocity is constant so the radial particles (or observers) are traveling like in an empty flat space with a constant velocity.

The fact that both timelike and null radial geodesics behave like in a flat space might lead to doubts whether the divergence of curvature scalars in the vicinity of $r=\chi_0$ (see (\ref{RicciScalar})) is accompanied by the main criterion for singularity which is geodesic incompleteness. However, inspection of the line element (\ref{static-metric}) immediately shows that the geometry cannot be extended beyond $r=\chi_0$ as a Lorentzian manifold since the metric signature changes there. Moreover, this point can be reached in finite proper time for radial timelike geodesic starting at finite radius $r_0>\chi_0$ (e.g. consider geodesic given by $t=\sqrt{(r_0-\chi_0)^2+1}\,\tau,\ r=-(r_0-\chi_0)\,\tau+r_0,\ \theta=const.,\ \phi=const.$ which terminates at $r=\chi_0$ at proper time $\tau=1$). The overall structure of the spacetime is captured on the Penrose-Carter diagram shown in Figure \ref{Penrose-diag} (adapted from \cite{mastersthesis}).

\begin{figure}
\centering
\begin{tikzpicture}[scale=0.9]

\coordinate (Origin) at (0,0);

\path  
	(Origin)
		+(90:4) coordinate[label=90:$i^+$] (ip)
		+(-90:4) coordinate[label=-90:$i^-$] (im)
		+(0:4) coordinate[label=0:$i^0$] (io) ;

\draw [thick]
	(im) --
		node[midway, below right] {$\cal{J}^-$}    
	(io) --  
		node[midway, above right] {$\cal{J}^+$}
	(ip);

\fill (ip) circle[radius=2pt];
\fill (io) circle[radius=2pt];
\fill (im) circle[radius=2pt];
	
\draw [dotted]
	(ip) --
		node[rotate=90, left=4pt, anchor=south, inner sep=4pt] {$r=0$}
	(im); 

\draw [dotted] 
	(1.0,0) -- 
		node[midway, above=-2pt] {$t=0$} 
	(io);


\draw [decorate,decoration=snake, thick]
	(ip)-- (0.2,3) -- (0.5,2) -- (0.75,1.5) -- (1.0,0.5) -- (1.0,0.0)  
		node[rotate=90, left=1pt, anchor=south, inner sep=7pt] {$r=\chi_0$} --
	(1.0,-0.5) -- (0.75,-1.5) -- (0.5,-2)-- (0.2,-3) -- (im);

\fill [decoration={snake}]
        [color=gray,opacity=0.2] (io) -- (ip)
          decorate { (ip)-- (0.2,3) -- (0.5,2) -- (0.75,1.5) -- (1.0,0.5) -- (1.0,0.0) 
          -- (1.0,-0.5) -- (0.75,-1.5) -- (0.5,-2)-- (0.2,-3) -- (im)} -- (io);

\draw[dotted, domain=0.22:1,smooth,variable=\u] plot ({0.0 +4.0*\u},{0.9 +0.9*cos(3.1416*\u r)});
\draw[dotted, domain=0.22:1,smooth,variable=\u] plot ({0.0 +4.0*\u},{-0.9 -0.9*cos(3.1416*\u r)});

\draw[dotted, domain=-1:1,smooth,variable=\u] plot ({1.4 +1.4*cos(3.1416*\u r)},{0.0 +4.0*\u});
\draw[dotted, domain=-1:1,smooth,variable=\u] plot ({0.9 +0.9*cos(3.1416*\u r)},{0.0 +4.0*\u});





\end{tikzpicture}
\caption{Penrose-Carter diagram of the metric $\d s^2 = \d {t}^2-\d {r}^2- \left({r}^2-{\chi_{0}^2}\right) \d{\Omega}^2$ which is being studied in this paper. The undulating line represents our singularity and the spacetime is asymptotically flat thus reaching to future $\cal{J}^+$ and past $\cal{J}^-$ null infinities.                        
}
\label{Penrose-diag}
\end{figure}

\section{Relation of area and acceleration}\label{appendixb}
Since we study a pair of neighboring geodesics, we can associate to it a timelike surface swept by the connecting vector $\mathbf{\delta}$ (see the text before equation (\ref{geodesic-deviation})) as we move along the radial geodesic. Since the acceleration is not constant in our case we will consider only short interval of proper time where it can be considered approximately constant. Moreover, since the neighboring (nonradial) geodesic can only be expressed by formulas involving elliptic integrals of the second kind we cannot proceed analytically.  Nevertheless, we can estimate the area swept by the proper distance $l(\tau)\approx\int_{0}^{\tilde{\varphi}(\tau)}\sqrt{r^2-\chi_{0}^{2}}\ \d \varphi$ between the reference radial geodesic (with $\varphi=0$) and the neighboring one (described by $\varphi=\tilde{\varphi}(\tau)$) over a small interval of the proper time $(0,\Delta\tau)$ using the equation (\ref{Elb}) to get
\begin{equation}
 A=\int_{0}^{\Delta\tau} l(\tau)\ \d \tau \stackrel{\Delta\tau \ll 1}{\approx} \sqrt{r^2-\chi_{0}^{2}}\ \tilde{\varphi}(0)\Delta\tau + O(\Delta\tau^{2})\ .
\end{equation}
Clearly the factor $\sqrt{r^2-\chi_{0}^{2}}$ (crucially appearing with opposite order of the power in the above equation and in (\ref{acceleration})) can be expressed using acceleration (\ref{acceleration}). Thus one can again obtain measure of acceleration expressed by area of timelike surface and the minimal area result shown in \cite{Rovelli:2013osa} for this kind of surface leads to an upper bound on acceleration.

\bibliographystyle{elsarticle-num}
\bibliography{robtrautall}

\begin{thebibliography}{10}
\expandafter\ifx\csname url\endcsname\relax
  \def\url#1{\texttt{#1}}\fi
\expandafter\ifx\csname urlprefix\endcsname\relax\def\urlprefix{URL }\fi
\expandafter\ifx\csname href\endcsname\relax
  \def\href#1#2{#2} \def\path#1{#1}\fi

\bibitem{Chase1970}
J.~E. Chase, \href{https://doi.org/10.1007/BF01646635}{Event horizons in static
  scalar-vacuum space-times}, Communications in Mathematical Physics 19~(4)
  (1970) 276--288.
\newblock \href {https://doi.org/10.1007/BF01646635}
  {\path{doi:10.1007/BF01646635}}.
\newline\urlprefix\url{https://doi.org/10.1007/BF01646635}

\bibitem{Tafel:2011aa}
J.~Tafel, {Static spherically symmetric black holes with scalar field}, Gen.
  Rel. Grav. 46 (2014) 1645, [Gen. Rel. Grav.46,1645(2014)].
\newblock \href {http://arxiv.org/abs/1112.2687} {\path{arXiv:1112.2687}},
  \href {https://doi.org/10.1007/s10714-013-1645-3}
  {\path{doi:10.1007/s10714-013-1645-3}}.

\bibitem{Tahamtan:2015sra}
T.~Tahamtan, O.~Svitek, {Robinson-Trautman solution with scalar hair}, Phys.
  Rev. D91~(10) (2015) 104032.
\newblock \href {http://arxiv.org/abs/1503.09080} {\path{arXiv:1503.09080}},
  \href {https://doi.org/10.1103/PhysRevD.91.104032}
  {\path{doi:10.1103/PhysRevD.91.104032}}.

\bibitem{Tahamtan:2016fur}
T.~Tahamtan, O.~Svitek, {Properties of Robinson--Trautman solution with scalar
  hair}, Phys. Rev. D94~(6) (2016) 064031.
\newblock \href {http://arxiv.org/abs/1603.07281} {\path{arXiv:1603.07281}},
  \href {https://doi.org/10.1103/PhysRevD.94.064031}
  {\path{doi:10.1103/PhysRevD.94.064031}}.

\bibitem{Janis:1968zz}
A.~I. Janis, E.~T. Newman, J.~Winicour, {Reality of the Schwarzschild
  Singularity}, Phys. Rev. Lett. 20 (1968) 878--880.
\newblock \href {https://doi.org/10.1103/PhysRevLett.20.878}
  {\path{doi:10.1103/PhysRevLett.20.878}}.

\bibitem{Wyman:1981bd}
M.~Wyman, {Static Spherically Symmetric Scalar Fields in General Relativity},
  Phys. Rev. D24 (1981) 839--841.
\newblock \href {https://doi.org/10.1103/PhysRevD.24.839}
  {\path{doi:10.1103/PhysRevD.24.839}}.

\bibitem{Fisher:1948yn}
I.~Z. Fisher, {Scalar mesostatic field with regard for gravitational effects},
  Zh. Eksp. Teor. Fiz. 18 (1948) 636--640.
\newblock \href {http://arxiv.org/abs/gr-qc/9911008}
  {\path{arXiv:gr-qc/9911008}}.

\bibitem{Janis:1970kn}
A.~I. Janis, D.~C. Robinson, J.~Winicour, {Comments on einstein scalar
  solutions}, Phys. Rev. 186 (1969) 1729--1731.
\newblock \href {https://doi.org/10.1103/PhysRev.186.1729}
  {\path{doi:10.1103/PhysRev.186.1729}}.

\bibitem{Wald:1980jn}
R.~M. Wald, {Dynamics in Nonglobally Hyperbolic, Static Space-times}, J. Math.
  Phys. 21 (1980) 2802--2805.
\newblock \href {https://doi.org/10.1063/1.524403}
  {\path{doi:10.1063/1.524403}}.

\bibitem{Horowitz:1995gi}
G.~T. Horowitz, D.~Marolf, {Quantum probes of space-time singularities}, Phys.
  Rev. D52 (1995) 5670--5675.
\newblock \href {http://arxiv.org/abs/gr-qc/9504028}
  {\path{arXiv:gr-qc/9504028}}, \href
  {https://doi.org/10.1103/PhysRevD.52.5670}
  {\path{doi:10.1103/PhysRevD.52.5670}}.

\bibitem{Ishibashi:1999vw}
A.~Ishibashi, A.~Hosoya, {Who's afraid of naked singularities? Probing timelike
  singularities with finite energy waves}, Phys. Rev. D60 (1999) 104028.
\newblock \href {http://arxiv.org/abs/gr-qc/9907009}
  {\path{arXiv:gr-qc/9907009}}, \href
  {https://doi.org/10.1103/PhysRevD.60.104028}
  {\path{doi:10.1103/PhysRevD.60.104028}}.

\bibitem{Konkowski:2001th}
D.~A. Konkowski, T.~M. Helliwell, {Quantum singularity of quasiregular
  space-times}, Gen. Rel. Grav. 33 (2001) 1131--1136.
\newblock \href {https://doi.org/10.1023/A:1010288501093}
  {\path{doi:10.1023/A:1010288501093}}.

\bibitem{Helliwell:2003tx}
T.~M. Helliwell, D.~A. Konkowski, V.~Arndt, {Quantum singularity in
  quasiregular space-times, as indicated by Klein-Gordon, Maxwell and Dirac
  fields}, Gen. Rel. Grav. 35 (2003) 79--96.
\newblock \href {https://doi.org/10.1023/A:1021307012363}
  {\path{doi:10.1023/A:1021307012363}}.

\bibitem{Konkowski:2011zz}
D.~A. Konkowski, T.~M. Helliwell, {Quantum singularities in static and
  conformally static space-times}, Int. J. Mod. Phys. A26 (2011) 3878--3888,
  [Int. J. Mod. Phys. Conf. Ser.03,364(2011)].
\newblock \href {http://arxiv.org/abs/1112.5488} {\path{arXiv:1112.5488}},
  \href {https://doi.org/10.1142/S2010194511001462, 10.1142/S0217751X11054334}
  {\path{doi:10.1142/S2010194511001462, 10.1142/S0217751X11054334}}.

\bibitem{Pitelli:2007jp}
P.~M. Pitelli, P.~S. Letelier, {Quantum Singularities in Spacetimes with
  Spherical and Cylindrical Topological Defects}, J. Math. Phys. 48 (2007)
  092501.
\newblock \href {http://arxiv.org/abs/0708.2052} {\path{arXiv:0708.2052}},
  \href {https://doi.org/10.1063/1.2779952} {\path{doi:10.1063/1.2779952}}.

\bibitem{Pitelli:2008pa}
J.~P.~M. Pitelli, P.~S. Letelier, {Quantum singularities in the BTZ spacetime},
  Phys. Rev. D77 (2008) 124030.
\newblock \href {http://arxiv.org/abs/0805.3926} {\path{arXiv:0805.3926}},
  \href {https://doi.org/10.1103/PhysRevD.77.124030}
  {\path{doi:10.1103/PhysRevD.77.124030}}.

\bibitem{Pitelli:2009kd}
J.~P.~M. Pitelli, P.~S. Letelier, {Quantum Singularities Around a Global
  Monopole}, Phys. Rev. D80 (2009) 104035.
\newblock \href {http://arxiv.org/abs/0911.2626} {\path{arXiv:0911.2626}},
  \href {https://doi.org/10.1103/PhysRevD.80.104035}
  {\path{doi:10.1103/PhysRevD.80.104035}}.

\bibitem{Tahamtan:2012sq}
T.~Tahamtan, O.~Gurtug, {Quantum singularities in a model of f(R) Gravity},
  Eur. Phys. J. C72 (2012) 2091.
\newblock \href {http://arxiv.org/abs/1205.5125} {\path{arXiv:1205.5125}},
  \href {https://doi.org/10.1140/epjc/s10052-012-2091-1}
  {\path{doi:10.1140/epjc/s10052-012-2091-1}}.

\bibitem{Tahamtan:2013vza}
T.~Tahamtan, O.~Sv{i}tek, {Resolution of curvature singularities from quantum
  mechanical and loop perspective}, Eur. Phys. J. C74~(8) (2014) 2987.
\newblock \href {http://arxiv.org/abs/1312.7806} {\path{arXiv:1312.7806}},
  \href {https://doi.org/10.1140/epjc/s10052-014-2987-z}
  {\path{doi:10.1140/epjc/s10052-014-2987-z}}.

\bibitem{Bronstein2012}
M.~P. Bronstein, \href{https://doi.org/10.1007/s10714-011-1285-4}{Republication
  of: Quantum theory of weak gravitational fields}, General Relativity and
  Gravitation 44~(1) (2012) 267--283.
\newblock \href {https://doi.org/10.1007/s10714-011-1285-4}
  {\path{doi:10.1007/s10714-011-1285-4}}.
\newline\urlprefix\url{https://doi.org/10.1007/s10714-011-1285-4}

\bibitem{Bronstein1936}
M.~P. {Bronstein}, {Quantization of gravitational waves}, Zhurnal
  Eksperimentalnoi i Teoreticheskoi Fiziki 6 (1936) 195.

\bibitem{Rovelli:2014ssa}
C.~Rovelli, F.~Vidotto, {Covariant Loop Quantum Gravity}, Cambridge Monographs
  on Mathematical Physics, Cambridge University Press, 2014.

\bibitem{Kuchar:1982eb}
K.~Kuchar, {Conditional symmetries in parametrized field theories}, J. Math.
  Phys. 23 (1982) 1647--1661.
\newblock \href {https://doi.org/10.1063/1.525550}
  {\path{doi:10.1063/1.525550}}.

\bibitem{Christodoulakis:2012eg}
T.~Christodoulakis, N.~Dimakis, P.~A. Terzis, G.~Doulis, T.~Grammenos,
  E.~Melas, A.~Spanou, {Conditional Symmetries and the Canonical Quantization
  of Constrained Minisuperspace Actions: the Schwarzschild case}, J. Geom.
  Phys. 71 (2013) 127--138.
\newblock \href {http://arxiv.org/abs/1208.0462} {\path{arXiv:1208.0462}},
  \href {https://doi.org/10.1016/j.geomphys.2013.04.009}
  {\path{doi:10.1016/j.geomphys.2013.04.009}}.

\bibitem{Christodoulakis:2013sya}
T.~Christodoulakis, N.~Dimakis, P.~A. Terzis, B.~Vakili, E.~Melas, et~al.,
  {Minisuperspace canonical quantization of the Reissner-Nordstrom black hole
  via conditional symmetries}, Phys. Rev. D 89~(4) (2014) 044031.
\newblock \href {http://arxiv.org/abs/1309.6106} {\path{arXiv:1309.6106}},
  \href {https://doi.org/10.1103/PhysRevD.89.044031}
  {\path{doi:10.1103/PhysRevD.89.044031}}.

\bibitem{Christodoulakis:2013xha}
T.~Christodoulakis, N.~Dimakis, P.~A. Terzis, {Lie point and variational
  symmetries in minisuperspace Einstein gravity}, J. Phys. A 47 (2014) 095202.
\newblock \href {http://arxiv.org/abs/1304.4359} {\path{arXiv:1304.4359}},
  \href {https://doi.org/10.1088/1751-8113/47/9/095202}
  {\path{doi:10.1088/1751-8113/47/9/095202}}.

\bibitem{Zampeli:2015ojr}
A.~Zampeli, T.~Pailas, P.~A. Terzis, T.~Christodoulakis, {Conditional
  symmetries in axisymmetric quantum cosmologies with scalar fields and the
  fate of the classical singularities}, JCAP 1605~(05) (2016) 066.
\newblock \href {http://arxiv.org/abs/1511.08382} {\path{arXiv:1511.08382}},
  \href {https://doi.org/10.1088/1475-7516/2016/05/066}
  {\path{doi:10.1088/1475-7516/2016/05/066}}.

\bibitem{Dimakis:2017qcf}
N.~Dimakis, A.~Karagiorgos, T.~Pailas, P.~A. Terzis, T.~Christodoulakis,
  {Discrete spectrum of the quantum Reissner-Nordstr{\"o}m geometry}, Phys.
  Rev. D95~(8) (2017) 086016.
\newblock \href {http://arxiv.org/abs/1703.05292} {\path{arXiv:1703.05292}},
  \href {https://doi.org/10.1103/PhysRevD.95.086016}
  {\path{doi:10.1103/PhysRevD.95.086016}}.

\bibitem{Karagiorgos:2017nta}
A.~Karagiorgos, T.~Pailas, N.~Dimakis, P.~A. Terzis, T.~Christodoulakis,
  {Quantum cosmology of a Bianchi III LRS geometry coupled to a source free
  electromagnetic field}, JCAP 1803~(03) (2018) 030.
\newblock \href {http://arxiv.org/abs/1710.02032} {\path{arXiv:1710.02032}},
  \href {https://doi.org/10.1088/1475-7516/2018/03/030}
  {\path{doi:10.1088/1475-7516/2018/03/030}}.

\bibitem{sym10030070}
T.~Christodoulakis, A.~Karagiorgos, A.~Zampeli,
  \href{http://www.mdpi.com/2073-8994/10/3/70}{Symmetries in classical and
  quantum treatment of einstein's cosmological equations and mini-superspace
  actions}, Symmetry 10~(3) (2018).
\newblock \href {https://doi.org/10.3390/sym10030070}
  {\path{doi:10.3390/sym10030070}}.
\newline\urlprefix\url{http://www.mdpi.com/2073-8994/10/3/70}

\bibitem{Perez:2012wv}
A.~Perez, {The Spin Foam Approach to Quantum Gravity}, Living Rev. Rel. 16
  (2013) 3.
\newblock \href {http://arxiv.org/abs/1205.2019} {\path{arXiv:1205.2019}},
  \href {https://doi.org/10.12942/lrr-2013-3} {\path{doi:10.12942/lrr-2013-3}}.

\bibitem{Thiemann:2007pyv}
T.~Thiemann, {Modern Canonical Quantum General Relativity}, Cambridge
  Monographs on Mathematical Physics, Cambridge University Press, 2007.
\newblock \href {https://doi.org/10.1017/CBO9780511755682}
  {\path{doi:10.1017/CBO9780511755682}}.

\bibitem{Rovelli:1994ge}
C.~Rovelli, L.~Smolin, {Discreteness of area and volume in quantum gravity},
  Nucl. Phys. B442 (1995) 593--622, [Erratum: Nucl. Phys.B456,753(1995)].
\newblock \href {http://arxiv.org/abs/gr-qc/9411005}
  {\path{arXiv:gr-qc/9411005}}, \href
  {https://doi.org/10.1016/0550-3213(95)00150-Q, 10.1016/0550-3213(95)00550-5}
  {\path{doi:10.1016/0550-3213(95)00150-Q, 10.1016/0550-3213(95)00550-5}}.

\bibitem{Ashtekar:1996eg}
A.~Ashtekar, J.~Lewandowski, {Quantum theory of geometry. 1: Area operators},
  Class. Quant. Grav. 14 (1997) A55--A82.
\newblock \href {http://arxiv.org/abs/gr-qc/9602046}
  {\path{arXiv:gr-qc/9602046}}, \href
  {https://doi.org/10.1088/0264-9381/14/1A/006}
  {\path{doi:10.1088/0264-9381/14/1A/006}}.

\bibitem{Reed:1975uy}
M.~Reed, B.~Simon, {Methods of Modern Mathematical Physics. 2. Fourier
  Analysis, Selfadjointness}, Academic, New York, 1975.

\bibitem{ronveaux1995heun}
A.~Ronveaux, F.~Arscott,
  \href{https://books.google.cz/books?id=5p65FD8caCgC}{Heun's Differential
  Equations}, Oxford science publications, Oxford University Press, 1995.
\newline\urlprefix\url{https://books.google.cz/books?id=5p65FD8caCgC}

\bibitem{Penrose:1986ca}
R.~Penrose, W.~Rindler, {Spinors and Space-time. Vol. 2: Spinor and Twistor
  Methods in Space-time Geometry}, Cambridge University Press, 1988.

\bibitem{Chandrasekhar:1985kt}
S.~Chandrasekhar, {The mathematical theory of black holes}, {Oxford, UK:
  Clarendon}, 1998.

\bibitem{Christodoulakis:2001mg}
T.~Christodoulakis, E.~Korfiatis, G.~Papadopoulos, {Automorphism inducing
  diffeomorphisms and invariant characterization of Bianchi type geometries},
  Commun.Math.Phys. 226 (2002) 377--391.
\newblock \href {http://arxiv.org/abs/gr-qc/0107050}
  {\path{arXiv:gr-qc/0107050}}, \href {https://doi.org/10.1007/s002200200611}
  {\path{doi:10.1007/s002200200611}}.

\bibitem{Zampeli2019}
A.~Zampeli, Minisuperspace quantisation via conditional symmetries, in:
  S.~Cacciatori, B.~G{\"u}neysu, S.~Pigola (Eds.), Einstein Equations: Physical
  and Mathematical Aspects of General Relativity: Domoschool 2018, Springer
  International Publishing, Birkh{\"a}user Cham, 2019, pp. 345--357.
\newblock \href {https://doi.org/https://doi.org/10.1007/978-3-030-18061-4_13}
  {\path{doi:https://doi.org/10.1007/978-3-030-18061-4_13}}.

\bibitem{Christodoulakis:1984gp}
T.~Christodoulakis, J.~Zanelli, {Operator Ordering in Quantum Mechanics and
  Quantum Gravity}, Nuovo Cim. B 93 (1986) 1--21.
\newblock \href {https://doi.org/10.1007/BF02728299}
  {\path{doi:10.1007/BF02728299}}.

\bibitem{Hartle_1987}
J.~B. Hartle, \href{http://dx.doi.org/10.1007/978-1-4613-1897-2_12}{Prediction
  in quantum cosmology}, NATO ASI Series (1987) 329--360\href
  {https://doi.org/10.1007/978-1-4613-1897-2_12}
  {\path{doi:10.1007/978-1-4613-1897-2_12}}.
\newline\urlprefix\url{http://dx.doi.org/10.1007/978-1-4613-1897-2_12}

\bibitem{Bohm:1951xw}
D.~Bohm, {A Suggested interpretation of the quantum theory in terms of hidden
  variables. 1.}, Phys.Rev. 85 (1952) 166--179.
\newblock \href {https://doi.org/10.1103/PhysRev.85.166}
  {\path{doi:10.1103/PhysRev.85.166}}.

\bibitem{Bohm:1951xx}
D.~Bohm, {A Suggested interpretation of the quantum theory in terms of hidden
  variables. 2.}, Phys.Rev. 85 (1952) 180--193.
\newblock \href {https://doi.org/10.1103/PhysRev.85.180}
  {\path{doi:10.1103/PhysRev.85.180}}.

\bibitem{Rovelli:2013osa}
C.~Rovelli, F.~Vidotto, {Evidence for Maximal Acceleration and Singularity
  Resolution in Covariant Loop Quantum Gravity}, Phys. Rev. Lett. 111 (2013)
  091303.
\newblock \href {http://arxiv.org/abs/1307.3228} {\path{arXiv:1307.3228}},
  \href {https://doi.org/10.1103/PhysRevLett.111.091303}
  {\path{doi:10.1103/PhysRevLett.111.091303}}.

\bibitem{Magueijo:2001cr}
J.~Magueijo, L.~Smolin, {Lorentz invariance with an invariant energy scale},
  Phys. Rev. Lett. 88 (2002) 190403.
\newblock \href {http://arxiv.org/abs/hep-th/0112090}
  {\path{arXiv:hep-th/0112090}}, \href
  {https://doi.org/10.1103/PhysRevLett.88.190403}
  {\path{doi:10.1103/PhysRevLett.88.190403}}.

\bibitem{Svitek:2014fga}
O.~Svitek, T.~Tahamtan, {Ultrarelativistic boost with scalar field}, Gen. Rel.
  Grav. 48~(2) (2016) 22.
\newblock \href {http://arxiv.org/abs/1406.6334} {\path{arXiv:1406.6334}},
  \href {https://doi.org/10.1007/s10714-016-2021-x}
  {\path{doi:10.1007/s10714-016-2021-x}}.

\bibitem{mastersthesis}
J.~\v{C}ern\'{y}, Canonical quantization of midisuperspace models, Master's
  thesis, Charles University, Prague (9 2018).

\end{thebibliography}
\end{document}